\documentclass[11pt]{article}
\usepackage{epic}

\usepackage{amsmath,amssymb}
\usepackage{amsthm}
\usepackage{amsfonts}
\usepackage{amscd}
\usepackage{mathrsfs}

\newcommand{\E}{{\cal E}}
\newcommand{\Sp}{{\cal S}}
\newcommand{\J}{{\cal J}}

\newcommand{\ads}{AdS_5\times S^5}
\newcommand{\sxx}{\nonumber \\ && }

\allowdisplaybreaks

\title{Reciprocity of higher conserved charges in the 
  $\mathfrak{sl}(2)$ sector of ${\cal N}=4$ SYM}

\author{Guido Macorini and Matteo Beccaria\\
Physics Department, Salento University and INFN, 73100 Lecce, Italy \\
E-mail:  macorini@le.infn.it, beccaria@le.infn.it}

\begin{document}
\maketitle

\begin{center}
\begin{minipage}{0.8\textwidth}
\centerline{\bf Abstract}
\medskip
We extend the anlysis of the generalized Gribov-Lipatov reciprocity to the 
higher conserved charges of type IIB superstring on $\ads$. The property is shown to  hold
for  twist $L=2$, and $3$ operators in the $\mathfrak{sl}(2)$ subsector.
\end{minipage}
\end{center}

%\keywords{ }

%\bodymatter

\section{Introduction and Discussion}
\label{sec:intro}

In the last years, integrability emerged as a
powerful tool in the investigation of the AdS/CFT correspondence.
The integrable spin chain description of the dilatation operator 
led to the  all-loop conjectured Bethe Ansatz  equations for the
$\mathfrak{psu}(2,2|4)$ 
algebra~\cite{Beisert:2003jj,Beisert:2003yb,Beisert:2005fw}, completely describing 
(once supplemented with the dressing phase~\cite{Beisert:2006ez,Beisert:2005wv}) the anomalous
dimensions (and the full tower of conserved charges) 
of the model, up to wrapping effects. 
The presence of an infinite set of
conserved charges $q_k$, forcing  the factorizability 
of the scattering matrix for elementary
excitations, is indeed a key manifestation 
of the integrability of a quantum model.\\

On the string side of the duality,  the corresponding
(classical) $\sigma-$model living on the string worldsheet in $AdS_5 \times
S^5$ is also integrable, and the tower of non local conserved charges was
derived in~\cite{Bena:2003wd,Mandal:2002fs,Engquist:2004bx}.\\   
 
Despite the physical relevance of $q_2$, identified with the
anomalous scaling dimension - string energy,  the first charge  does not play a special role 
from the point of view of the
integrability, and all the
charges are on equal footing. In~\cite{Arutyunov:2003rg} the weak-strong coupling
correspondence of the full tower of charges in the $\mathfrak{su}(2)$ sector
has been studied, but  the physical meaning and properties of the higher
conserved charges remains less understood.\\

In this work we  investigate the reciprocity properties
(see~\cite{Beccaria:2010tb} for a review) 
of the first higher conserved charges in the $\mathfrak{sl}(2)$ sector.
Reciprocity has its far origin in QCD in a symmetric treatment of the 
Deep Inelastic Scattering (DIS) and electron-positron into hadrons.
The modified symmetric DGLAP kernel  $P(N)$ in the evolution equation obeys
the relation: $  \gamma(N) = P\left(N+\textstyle{\frac{1}{2}\gamma(N)}\right)$,
where $\gamma(N)$ is the lowest anomalous dimension, and the reciprocity
can be recast in the form of an asymptotic, large spin condition $ P(N) =
\sum_{\ell\ge 0} \frac{a_\ell(\log\,J^2)}{J^{2\,\ell}},$ where $J^2 = N\,(N+1)$,
$a_\ell$ are coupling-dependent polynomials and 
$J^2$ is the Casimir of the collinear subgroup $SL(2,\mathbb{R})\subset
SO(2,4)$. This condition can also be interpreted as a parity invariance $J
\to -J$ in the large spin regime.\\  
    
The $\mathfrak{sl}(2)$ sector, spanned by single trace operators  
$\mathcal{O} \sim {\rm Tr}\,(\mathcal{D}^{n_1} Z \dots \mathcal{D}^{n_L} Z),$
is a closed subsector of the theory 
under perturbative renormalization;
$N=\sum n_i$ is the total spin and $L$ is the classical dimension minus the spin (twist) 
of the operator. The relevant dual string state is the classical folded $(S,J)$ string solution,
describing a string extended in
the radial direction of $AdS_5$ and rotating in
$AdS_5$, with center of mass moving on a circle of $S^5$~\cite{Frolov:2002av,folded}. This
solution is linked via analytic continuation $ E \to -J_1,\; S \to
J_2,\; J \to  -E$ to the string configuration $(J_1, J_2)$ with two 
angular momenta on $S^5$, for which the higher charges 
at strong coupling have been constructed in~\cite{Arutyunov:2003rg}
by using  the B\"acklund transformations of the 
integrable classical string $\sigma$-model.\\ 

Analysing the first two  charges at weak coupling $q_{4,6}$ (respectively at
3-loop plus 4-loop dressing part and 2-loop level) and the first 
charges at strong coupling at classical level, we find that the reciprocity condition can
be consistently generalized for all the tested higher charges.

\section{Higher Charges and Reciprocity at Weak Coupling}
\label{sec:weak}

In the weak coupling regime the closed formulae for multi-loops higher charges
can be efficiently computed following the Baxter approach~\cite{Bax72,Derkachov:2002tf,KorTrick2} toghether with the
maximum transcendentality Ansatz (and then completed by the dressing factor
starting from the four-loop order), resulting in a combination of harmonic
sums of definite transcendentality~\cite{Beccaria:2009yt}. 
The reciprocity condition for the full tower of conserved charges can be
generalized from the condition for $q_2$ 
defining the kernel $P_r(N)$ as 
\begin{equation}
  q_r(N) = P_r\left(N + \frac{1}{2}~ q_2(N)\right).
  % \label{eq:P}
\end{equation}
This equation emphasizes the role of the renormalized conformal spin, as also suggested by light cone
quantization. 
Reciprocity implies a constraint on the form of  the expansion of $P_r$ at
large $N$, which should involve  only integer inverse powers of
$N\,(N+1)$. The check of this property is easier after a rewriting of the charges  in
terms of the $\Omega$ basis~\cite{Beccaria:2009vt}, where the reciprocity simply means that
the $\Omega$ must have odd positive or even negative indices. We report here
only the first, parity respecting results for the higher charge $q_4$:\\

{\bf $L=2$, three-loops reciprocity of $q_4$}
{\small
  \begin{eqnarray}
    P_4^{(1)} &=&16 \left(\Omega _3+6 \Omega _{-2,1}\right), \\
    P_4^{(2)} &=& -\frac{16}{5} (\pi ^4 \Omega _1+120 \Omega _{-4,1}+20 \pi ^2
    \Omega _{-2,1}+60 \Omega _{-2,3}+60 \Omega _{1,-4}+20 \pi ^2 \Omega _{1,-2}
    \sxx+120 \Omega _{-2,1,-2} +120 \Omega _{1,-2,-2}-480 \Omega _{1,-2,1,1}), \\
    P_4^{(3)} &=& \frac{32}{15} (180 \zeta (3) \Omega _{-4}+2 \pi ^6 \Omega _1+3
    \pi ^4 \Omega _3-30 \pi ^2 \Omega _5-720 \Omega _7+900 \Omega _{-6,1}+240 \pi
    ^2 \Omega _{-4,1}
    \sxx +540 \Omega _{-4,3}+30 \pi ^4 \Omega _{-2,1}+60 \pi ^2 \Omega
    _{-2,3}+720\Omega _{1,-6}+240 \pi ^2 \Omega _{1,-4}+36 \pi ^4 \Omega _{1,-2}
    \sxx +180 \Omega _{3,-4} +60 \pi ^2 \Omega _{3,-2}-180 \Omega
    _{5,-2}+2520\Omega _{-4,-2,1}+2160 \Omega _{-4,1,-2}
    \sxx +1080 \Omega _{-2,-4,1}+360 \Omega _{-2,-2,3} + 1800 \Omega _{-2,1,-4} +
    120 \pi ^2 \Omega _{-2,1,-2}
    \sxx +1080 \Omega _{-2,3,-2}+1440 \Omega _{1,-4,-2}+2160 \Omega _{1,-2,-4}
    \sxx +240 \pi ^2 \Omega _{1,-2,-2}+1440 \Omega _{1,1,5}+2160
    \Omega_{1,5,1}+360 \Omega _{3,-2,-2}+720 \Omega _{5,1,1}
    \sxx -1440 \Omega _{-4,1,1,1}+2160 \Omega _{-2,-2,-2,1}
    \sxx +1440 \Omega _{-2,-2,1,-2}+720 \Omega _{-2,1,-2,-2}-2880 \Omega
    _{1,-4,1,1}
    \sxx +1440 \Omega _{1,-2,-2,-2}-960 \pi ^2 \Omega _{1,-2,1,1}-1440 \Omega
    _{1,-2,1,3}-1440 \Omega _{1,-2,3,1}-1440 \Omega _{1,1,-4,1}
    \sxx -960 \pi ^2 \Omega _{1,1,-2,1}-1440 \Omega _{1,1,-2,3}-1440 \Omega
    _{3,-2,1,1}-2880 \Omega _{-2,-2,1,1,1}
    \sxx -2880 \Omega _{-2,1,-2,1,1}-5760 \Omega _{-2,1,1,-2,1}-2880
    \Omega_{-2,1,1,1,-2}-2880 \Omega _{1,-2,-2,1,1}
    \sxx -5760 \Omega _{1,-2,1,-2,1} -5760 \Omega _{1,-2,1,1,-2}-11520
    \Omega_{1,1,-2,-2,1}
    \sxx -5760 \Omega _{1,1,-2,1,-2}+11520 \Omega _{1,1,-2,1,1,1} +360
    \Omega_{1,1} \zeta (5)-240 \pi ^2 \Omega _{1,1} \zeta (3)
    \sxx-720 \Omega _{-2,1,1} \zeta (3)-720 \Omega _{1,-2,1} \zeta (3) -720 \Omega
    _{1,1,-2} \zeta (3))
  \end{eqnarray}}

{\bf $L=2$, four-loops reciprocity of the dressing part of $q_4$}
{\small
  \begin{eqnarray}
    P_4^{(4, \rm dressing)} &=& 3072 \Omega _{-6}+3072 \Omega _{-2,-4}+3072 \Omega
    _{5,1}-18432 \Omega _{-4,1,1}
    \sxx -12288 \Omega _{-2,1,3}-12288 \Omega _{-2,3,1}-6144 \Omega _{1,-4,1}-6144
    \Omega _{1,-2,3}
    \sxx -24576 \Omega _{-2,-2,1,1}-12288 \Omega _{-2,1,-2,1}-12288 \Omega
    _{1,-2,-2,1}
    \sxx +98304 \Omega _{-2,1,1,1,1}+24576 \Omega _{1,-2,1,1,1}.
  \end{eqnarray}}

\section{Higher Charges and Reciprocity at Strong Coupling}
\label{sec:strong}

The string state dual of the gauge operators is the semiclassical
$\mathfrak{sl}(2)$ folded string. As anticipated, it is related to the $(J_1,
J_2)$ string by an analytic continuation, mapping one into another the
$\sigma$-models describing the strings on  $AdS_3 \times S^1$ and $R \times
S^3$, as well as the relative equations of motion, their solutions and the
conserved charges. 
Energy ${\cal E} = E/\sqrt{\lambda}$, spin ${\cal S} = S/\sqrt\lambda$,
and angular momentum  $\J = J/\sqrt\lambda $  for the folded string are related by
\begin{eqnarray}
  && \sqrt{\kappa^2-\J^2} =
  \frac{1}{\sqrt\eta}\,{}_2F_1\left(\frac{1}{2},
    \frac{1}{2}, 1, -\frac{1}{\eta}\right), \quad
  \omega^2-\J^2 = (1+\eta)\,(\kappa^2-\J^2),\\ 
  % \end{eqnarray}
  % \begin{eqnarray}
  %   \label{eq:spin1}
  && \Sp =
  \frac{\omega}{\sqrt{\kappa^2-\J^2}}\,\frac{1}{2\,\eta\,\sqrt\eta}\,{}_2F_1\left(\frac{1}{2},
    \frac{3}{2}, 2, -\frac{1}{\eta}\right), \quad
  % \label{eq:energy1}
  \E =
  \frac{\kappa}{\sqrt{\kappa^2-\J^2}}\,\frac{1}{\sqrt\eta}\,{}_2F_1\left(-\frac{1}{2},
    \frac{1}{2}, 1, -\frac{1}{\eta}\right)  
\end{eqnarray}
and for the comparison with the gauge theory results we are interested in the
slow string limit; using $\J$ as an expansion parameter, the quantum
contribution to the energy can be computed as:
\begin{eqnarray}
  \eta(\Sp, \J) = \eta^{(0)}(\Sp) + \eta^{(2)}(\Sp)\,\J^2 +
  \eta^{(4)}(\Sp)\,\J^4 + \cdots~,\\
  \Delta = \E -\Sp  = \Delta^{(0)}(\Sp) + \Delta^{(2)}(\Sp)\,\J^2 +
  \Delta^{(4)}(\Sp)\,\J^4 + \cdots~.
\end{eqnarray}
Introducing the function $f$ defined as $\Delta(\Sp) = {\cal E}({\cal S})-{\cal S} =
f\left({\cal S} + \frac{1}{2}\,{\cal E}({\cal S})\right)$
and expanding perturbatively, reciprocity is translated in the absence of inverse odd powers
of $\Sp$ in the expansions.\\

Higher charges ${\cal E}_{4,6,\dots}$ can be constructed in the $\mathfrak{su}(2)$ sector 
by using  the B\"acklund transformation method~\cite{Arutyunov:2003rg}, and
then analytically continued ($t\to -1/\eta, \; {\cal E}_2\to J$).  
As an example, for the first non-vanishing charge 
${\cal E}_4$ we get: 
\begin{eqnarray}
  & {\cal E}_4 = -\frac{16}{\pi^2\,{\cal E}_2}\,Z_1(t) +
  \frac{32}{\pi^4\,{\cal E}_2^3}\, Z_2(t),& \nonumber \\
  % \ee
  % where
  % \ba
  & Z_1(t) = \mathbb{K}(t)[\mathbb{E}(t)+(t-1)\mathbb{K}(t)],
  \qquad
  Z_2(t) = t(t-1) \mathbb{K}(t)^4 &
\end{eqnarray}
where $t$ is a modular parameter.
In analogy with the case of the energy, we propose to 
test reciprocity on the functions $f_k$ defined by 
\begin{equation}
  \label{eq:functional}
  Z_k({\cal S}) = f_k\left({\cal S} + \frac{1}{2}{\cal E}({\cal S})\right),\;\; Z_k({\cal S}) \equiv Z_k(-1/\eta({\cal S})).
\end{equation}
Using the  Lagrange-B\"urmann formula~\cite{Basso:2006nk}
\begin{equation}
  f({\cal S}) = \sum_{k=0}^\infty\frac{1}{k!}\left(\frac{d}{d{\cal
        S}}\right)^{k-1}\left[
    \left(-\frac{\Delta({\cal S})}{2}\right)^k\,Z'({\cal S})\right] = Z({\cal
    S})-\frac{1}{2}\,\Delta({\cal S})\,Z'({\cal S}) + \cdots
\end{equation}
from $\eta = \eta(\Sp, \J)$ we obtain  an expansion for
$ f_k(\Sp)  = f^{(0)}_k(\Sp) + f^{(2)}_k(\Sp)\,\J^2 + f^{(4)}_k(\Sp)\,\J^4 + \cdots~$
and computing  0-th order correction for $Z_1$ and $Z_2$ we find the result
{\small
  \begin{eqnarray}
    \label{eq:fz1}
    f_1^{(0)} &=& -\frac{1}{4} \left(\log \bar{\Sp}-2\right) \log \bar{\Sp} + 
    \mbox{\fbox{$\displaystyle 0\cdot\frac{1}{\bar{\Sp}}$}}
    % \sxx 
    +2 \left(2-3 \log \bar{\Sp}\right) \log \bar{\Sp} \,\frac{1}{\bar{\Sp}^2}
    +
    \mbox{\fbox{$\displaystyle 0\cdot\frac{1}{\bar{\Sp}^3}$}}
    \sxx +\left(80 \log
      ^3\bar{\Sp}-118 \log ^2\bar{\Sp}+23 \log \bar{\Sp}+1\right)
    \,\frac{1}{\bar{\Sp}^4}+
    \mbox{\fbox{$\displaystyle 0\cdot\frac{1}{\bar{\Sp}^5}$}}+\cdots, \\
    f_2^{(0)} &=& \frac{1}{16} \log ^4\bar{\Sp}+
    \mbox{\fbox{$\displaystyle 0\cdot\frac{1}{\bar{\Sp}}$}}
    % \sxx 
    +\log ^4\bar{\Sp} \,\frac{1}{\bar{\Sp}^2} +
    \mbox{\fbox{$\displaystyle 0\cdot\frac{1}{\bar{\Sp^3}}$}}
    \sxx -\frac{1}{2} \left(\log ^3\bar{\Sp} \left(16 \log ^2\bar{\Sp}-22 \log
        \bar{\Sp}-1\right)\right) \,\frac{1}{\bar{\Sp}^4}+
    \mbox{\fbox{$\displaystyle 0\cdot\frac{1}{\bar{\Sp^5}}$}}\,,
  \end{eqnarray}}
where the absence of inverse odd powers
of $\Sp$, highlighted by the boxes, clearly supports parity
invariance. The procedure can be straightforwardly extended to 
the next conserved charges, showing parity
invariance in all the tested cases.

\end{document}